**Chapter 1**

# High-*Q* Lithium Niobate Microcavities and Their Applications

Cheng Wang[*,§], Mian Zhang[†,¶] and Marko Lončar[‡,∥]

[*] *Department of Electrical Engineering, City University of Hong Kong, Kowloon, Hong Kong, China*
[†] *HyperLight Corporation, 501 Massachusetts Avenue, Cambridge, MA 02139, USA*
[‡] *John A. Paulson School of Engineering and Applied Sciences, Harvard University, Cambridge, MA 02138, USA*
[§] *cwang257@cityu.edu.hk*
[¶] *mian@hyperlightcorp.com*
[∥] *loncar@seas.harvard.edu*

Lithium niobate (LN) is an excellent nonlinear optical and electro-optic material that has found many applications in classical nonlinear optics, optical fiber communications and quantum photonics. Here we review the recent development of thin-film LN technology that has allowed the miniaturization of LN photonic devices and microcavities with ultrahigh quality factors. We discuss the design principle of LN devices that makes use of the largest nonlinear coefficients, various device fabrication approaches and resulting device performances, and the current and potential applications of LN microcavities.

## 1. Introduction

Since the discovery of optical nonlinearities in the 1960s, lithium niobate (LiNbO$_3$, or LN) has been the most widely used second-order ($\chi^{(2)}$) material, with applications ranging from nonlinear wavelength conversion for classical and quantum light source [1], optical modulators for data communications [2], as well as surface acoustic wave (SAW) based electronic components for mobile phone industry [3]. Compared with other common photonic materials, LN holds many favorable properties both in terms of nonlinear and linear optics. The largest $\chi^{(2)}$ tensor component in LN is aligned diagonally ($\chi^{(2)}_{zzz}$), and is large for both nonlinear wavelength conversion (known as $d_{33}$) and electro-optic modulation ($r_{33}$). As a linear optical material, LN possesses relatively high ordinary and extraordinary refractive indices ($n_o = 2.21$, $n_e = 2.14$, at 1550 nm), and is highly transparent in a wide wavelength range from 400 nm (blue) to 5 μm (mid-infrared), with an OH- absorption peak at 2.87 μm [4]. A comparison of important linear and nonlinear optical properties among representative $\chi^{(2)}$ materials is shown in Table 1.

Table 1. Optical properties of representative $\chi^{(2)}$ materials (~ 1500 nm wavelength)

| Material | Largest $d$ coefficient | Largest $r$ coefficient | Refractive index | Transparency window |
|---|---|---|---|---|
| LiNbO$_3$ [4] | $d_{33}$ = 27 pm/V | $r_{33}$ = 27 pm/V | ~ 2.2 | 400 nm – 5 μm |
| BBO [4] | $d_{22}$ = 1.9 pm/V | $r_{51}$ = 2.1 pm/V | ~ 1.6 | 200 nm – 2.6 μm |
| KTP [4] | $d_{33}$ = 14.6 pm/V | $r_{33}$ = 35 pm/V | ~ 1.8 | 350 nm – 4.5 μm |
| GaAs [4] | $d_{36}$ = 119 pm/V | $r_{41}$ = 1.53 pm/V | ~ 3.4 | 1.1 – 17 μm |
| AlN [5, 6] | $d_{33}$ = 4.7 ± 3 pm/V | $r_{33}$ ~ 1 pm/V | ~ 2.1 | 220 nm – 13.6 μm |
| GaN [7, 8] | $d_{33}$ ~ 20 pm/V | $r_{33}$ = 1.91 ± 0.35 | ~ 2.3 | 365 nm – 13.6 μm |
| SiC [9-11] | $d_{33}$ ~ 10 pm/V | $r_{33}$ = 2.7 ± 0.5 | ~ 2.5 | > 500 nm |

Since the $\chi^{(2)}$ nonlinear processes are much weaker than linear optical processes, it is a natural choice to use high-$Q$ microcavities to enhance the $\chi^{(2)}$ processes. However, standard photonic device fabrication techniques commonly used in other material systems are not readily compatible with LN, the challenges of which will be discussed in more details in Sections 3&4. Instead, optical waveguides in conventional LN nonlinear optical devices and electro-optic modulators are typically achieved by metal in-diffusion (e.g. Ti) or ion exchange (e.g. proton), which locally perturb the LN crystal and induces a small refractive index increase (Δn ~ 0.02) (Fig. 1) [12]. The low index contrast results in weak light confinement and large bending radii (~ 10 cm), preventing the realization of tightly confined microcavities in LN.

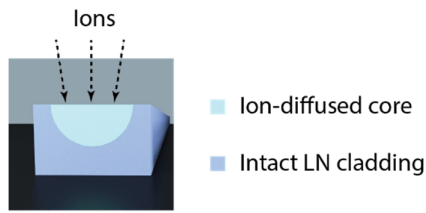

Fig. 1. Schematic of traditional ion-diffused waveguides. The low index contrast between waveguide core and cladding results in large mode areas and weak nonlinear interactions.

Where there is a will, there is a way. Or perhaps more than one way. Many different technological approaches have been extensively studied in recent years in order to realize high-$Q$ microcavities in LN. Along with such technology development, we see a continuous improvement in the $Q$ factors of these cavities, as well as the emergence of a variety of applications.

In this chapter, we provide an overview of basic design principles for LN devices, recent advances in LN fabrication technology, as well as their applications in nonlinear optics and electro-optics.

## 2. Design principles of lithium niobate devices

Needless to say, most applications of LN photonic devices are closely associated with $\chi^{(2)}$ processes. However, a rigorous mathematic treatment of $\chi^{(2)}$ processes could be quite complicated, since solving a three-wave-mixing process requires tensor analysis at the rank 3. The full $\chi^{(2)}$ tensor has a total of 27 components, connecting the 3 waves and a total of 9 vector field components. Taking sum-frequency generation ($\omega_1 + \omega_2 \rightarrow \omega_3$) as an example, the generated nonlinear polarization at $\omega_3$ can be expressed as:

$$P_i(\omega_3) = 2\epsilon_0 \sum_{jk} \chi^{(2)}_{ijk}(\omega_3, \omega_1, \omega_2) E_j(\omega_1) E_k(\omega_2) \quad (1)$$

where $i, j, k \in \{x, y, z\}$ [1].

Fortunately, in most LN applications such full treatment is not required. Instead, there is only one critical component out of the 27 – the $\chi^{(2)}_{zzz}$, since this component is much larger than all other ones [4]. Eq. (1) can therefore be reduced to a scalar equation. Note that different notations such as $d_{33}$ and $r_{33}$ are often used for certain processes, which we will explain next, but keep in mind that they all correspond to $\chi^{(2)}_{zzz}$.

*Long story short, there is only one design rule for almost all LN photonic devices: align the directions of all electric fields with the z axis.*

As shown in Fig. 2, taking a z-cut LN crystal as an example, transverse-magnetic (TM) optical modes are preferred for optical waves, and electrical contacts should be placed on top and bottom of the optical waveguide.

For second harmonic generation (SHG) [Fig. 2(a)], where pump light at a frequency of $\omega$ is frequency doubled to a signal frequency of $2\omega$, the generated nonlinear polarization at $2\omega$ can be expressed as:

$$P_z(2\omega) = 2\epsilon_0 d_{33} E_z^2(\omega) \quad (2)$$

where $d_{33} = \frac{1}{2}\chi^{(2)}_{zzz}$ is ~ 30 pm/V for LN and $E_z(\omega)$ is the pump electric field z component [1].

For electro-optic modulation [Fig. 2(b)], a DC (or microwave) electric field induces a change in optical index of refraction. For LN, the index change for z-polarized light can be expressed as:

$$\Delta n_z = \frac{1}{2} n_z^3 r_{33} E_z \qquad (3)$$

where $n_z$ is the linear refractive index for z-polarized light, $r_{33} = 30.9$ pm/V for LN, and $E_z$ is the DC (or microwave) electric field in z direction [1].

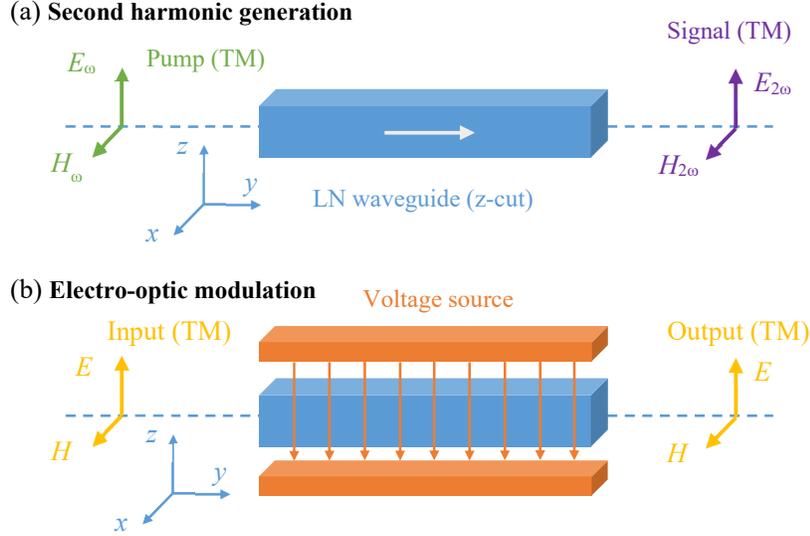

Fig. 2. Cross-section illustrations showing the ideal alignment of the electric fields with LN crystal (z-cut in this case) in order to utilize the $\chi^{(2)}_{zzz}$ components. (a) For SHG, TM modes are preferred. (b) For electro-optic modulation, TM optical mode and top-bottom electrodes are preferred.

## 3. Lithium niobate on insulator (LNOI)

Just like the majority of silicon photonics integrated circuits rely on silicon-on-insulator (SOI) substrates, a straightforward approach to build nanophotonic devices in LN is to use sub-micron thick, single-crystalline LN films that stand on top of a transparent and lower refractive index substrate, typically silica ($SiO_2$). These substrates are called lithium niobate on insulator (LNOI). The LNOI wafers are fabricated using a "Smart-cut" ion-slicing process (Fig. 3). First, $He^+$ ions with certain energies are implanted into a bulk LN substrate, creating a damage layer at a depth of interest [Fig. 3(a)]. The ion-implanted LN substrate is then flipped over and bonded on top of a desired substrate (e.g. thermal oxide on silicon) [Fig. 3(b)]. The bonded substrates are subsequently annealed to thermally induce stress in the damage layer, resulting in the splitting of LN substrate at the damage layer [Fig.

3(c)]. The resulting LNOI substrate is further polished to remove the residual damage layer and to create a smooth top surface, and annealed to restore the optical properties of LN in the thin film layer [Fig. 3(d)] [13, 14].

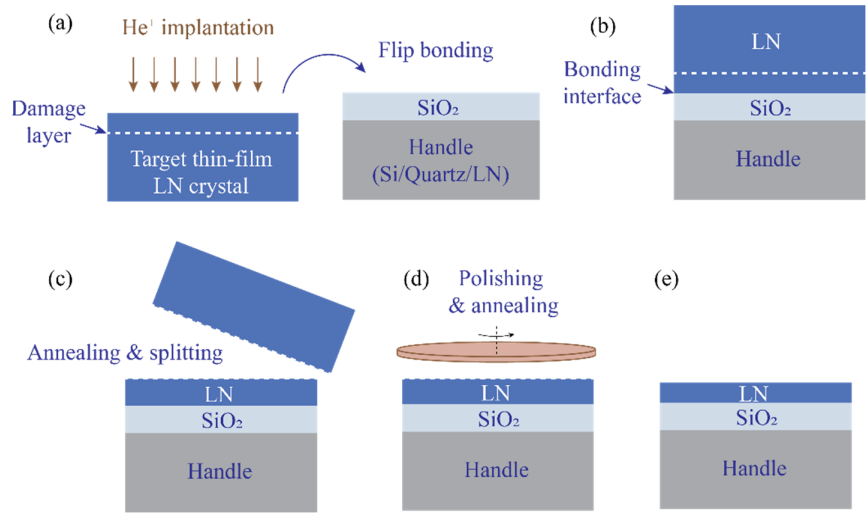

Fig. 3. Schematic of LNOI wafer production [13].

Thanks to the industrial development of LNOI technology in the past decade, high-quality LNOI wafers are now commercially available at sizes up to 4 inches, with various choices of crystal orientations and bottom handles. There are two major choices of LNOI crystal orientations – z-cut and x-cut. As we have discussed in Section 2, for z-cut substrates, transverse-magnetic (TM) optical modes and top/bottom electrodes are preferred, while for x-cut substrates, transverse-electric (TE) modes and in-plane electrodes are typically used. It is relatively easier to design devices for z-cut substrates, since the crystal is isotropic within the device layer (x-y plane). In this case, light does not experience changes in refractive index or nonlinear optic coefficients when it travels along different directions. However, z-cut LN crystal is known to experience more piezoelectric- and pyroelectric-induced electrical charge accumulation and requires special handling procedures during device fabrication [2]. For the same reason, modulators made of z-cut LN crystals are more prone to long-term bias drift issues [2]. Z-cut crystal requires top/bottom electrode structures, which could be inconvenient for microwave testing purposes. In contrast, x-cut crystals allow for in-plane electrode designs that can be flexibly defined using planar lithography approaches and are naturally compatible with standard microwave testing equipment (e.g. ground-signal-6ground probes). However, due to the anisotropy within the device layer, light only experiences the prominent nonlinear effect while traveling in one direction (y direction), therefore the nonlinear interaction could be compromised in a circular cavity. Racetrack-shaped microcavities are sometimes adopted to maximize the nonlinear effects [15].

The LNOI production processes also allow for different choices of bottom substrates (Fig. 4). Apart from the possibility of pre-embedding metal contact layer in the LNOI wafers, popular handle materials include LN, silicon and quartz. LN handles possess the same thermal expansion constants as the device layer, therefore provide the best wafer-bonding quality. Silicon handles offer the possibility of using thermally grown $SiO_2$ as the bottom cladding layer. Thermal $SiO_2$ possesses much lower optical absorption losses than $SiO_2$ grown on top of LN or quartz handles using plasma-enhanced chemical vapor deposition (PECVD). Quartz is an attractive handle material for applications that require high-performance microwave delivery (e.g. high-speed electro-optic modulators) due to its low microwave loss.

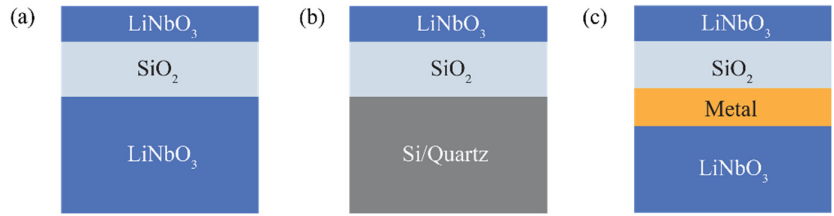

Fig. 4. A few examples of popular LNOI handle substrates. (a) Standard LNOI with LN handle substrate. (b) LNOI with silicon or quartz substrate. (c) LNOI with a pre-embedded bottom metal electrode layer.

It is worth mentioning that, although commercial LNOI technology is still in its early age and the substrates are more expensive than SOI wafers, we envision that further technology development and market expansion will drive the price down significantly in the foreseeable future. Bulk LN crystals are typically grown using Czochralski process similar to that used for silicon, and could be grown at > 6-inch scale. The LNOI production also follows a standard "Smart-cut" process. Therefore we believe the material property of LN is well suited for high-volume production of LNOI wafers and mass production of LN photonic devices.

## 4. Microcavity fabrication in LN

Although LNOI substrates are now commercially available, microstructuring of LN, in particular its dry etching, is still a challenging task. The fluorine-based plasma that is often used for dry etching produces non-volatile etching product (e.g. LiF), which could be redeposited on the etched surfaces and sidewalls, increasing surface roughness and reducing the etching rate [16].

Over the past decade, many creative fabrication techniques have been proposed and adopted in order to realize low-loss photonic structures and high-$Q$ microcavities in LN. Dry-etched microcavities with $Q$ factors on the order of $10^7$ [17-20] and mechanically polished cavities with $Q$ factors exceeding $10^8$ [21-23] have been reported. In this section, we try to provide a comprehensive review of the different fabrication strategies. Specifically, there are three major techniques: direct dry etching method, heterogeneous integration method, and rotatory mechanical polishing method.

### 4.1. *Direct dry etching of LN*

The most straightforward way of realizing highly confined photonic structures in the LNOI platform is to directly structure LN material using dry etching processes, similar to those used in other common photonic materials. Although LN is historically regarded as a difficult-to-etch, or even impossible-to-etch material, the past decade has witnessed an impressive development of LN dry etching technology, increasing the $Q$-factors of dry-etched LN resonators from ~ 4,000 in 2007 [24] to > 10,000,000 in 2018 [17-19].

Due to the difficulty in chemical etching of LN, most high-$Q$ LN photonic structures to date have been realized using physical etching approaches (e.g. Ar+ plasma or focused ion beams). Although such processes typically have low etching rates (~ 30 nm/min) and low etching selectivity (~ 1:1), they benefit from the shallow etching depths (< 500 nm) required in the LNOI platform.

The typical fabrication flow of a dry-etched LN resonator involves three major steps: lithography, dry etching and post-fabrication smoothening and undercutting (Fig. 5).

First, electron-beam lithography or photolithography tools are used to define the resonator patterns. Electron-beam lithography typically provides higher resolution required for fine structures like microrings and photonic crystals, while photolithography can be used to produce larger structures like microdisks at high throughput. Apart from these standard lithography tools, Y. Cheng's team at Chinese Academy of Sciences have achieved a range of high-quality LN resonators using femtosecond laser machining [19, 25-27]. This approach uses focused femtosecond laser pulses to directly remove the LN material through laser ablation, and could offer relatively high throughput and resolution.

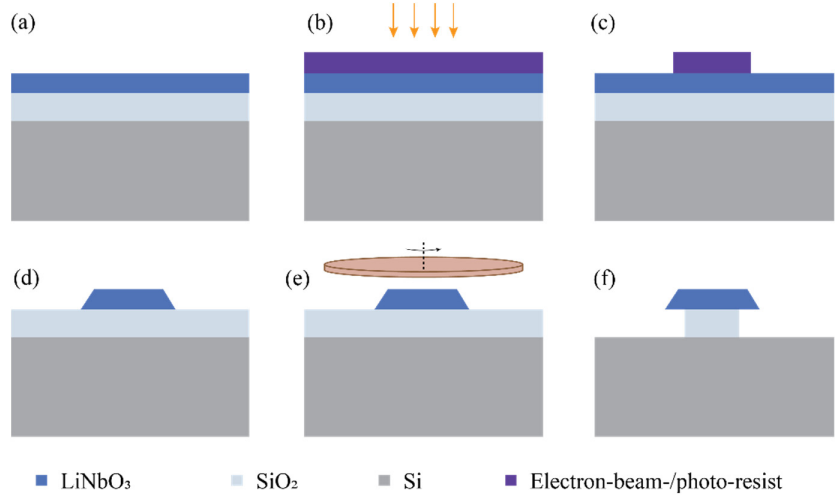

Fig. 5. Schematic illustration of typical fabrication procedures of dry etched LN devices. (a) An intact LNOI substrate is (b) spin-coated with electron-beam resist or photoresist. The resist is then exposed using electron-beam lithography or photolithography and (c) developed to define the device patterns. (d) Dry etching, typically in Ar+ plasma, is performed to transfer the resist patterns into LN device layer. (e) Post-fabrication smoothening techniques (e.g. mechanical polishing) could be applied to further reduce the sidewall roughness. (f) An optional hydrofluoric acid (HF) wet etching could be performed to create a suspended structure in case needed.

As we discussed above, the dry etching of LN is typically realized using Ar+-based physical bombardment, therefore the etching selectivity generally depends on the hardness of etching masks. Photoresists or electron-beam resists provide the highest sidewall smoothness since they are directly defined by lithography and can be reflowed to further improve the smoothness. However, at the same time, most of these materials are polymer based, resulting in relatively low etching selectivity. Metal masks provide the highest etching selectivity, but tend to have more sidewall roughness due to the existence of grains. Some relatively hard dielectric materials, e.g. $SiO_2$ and Si, are also popular etching masks that could provide a good balance between etching selectivity and mask smoothness. Due to the nature of physical etching processes, the etched LN sidewalls are often slanted with an angle to the vertical direction. Yet photonic structures with slanted sidewalls can support well-defined optical modes and are suitable for most applications.

**Microdisks**

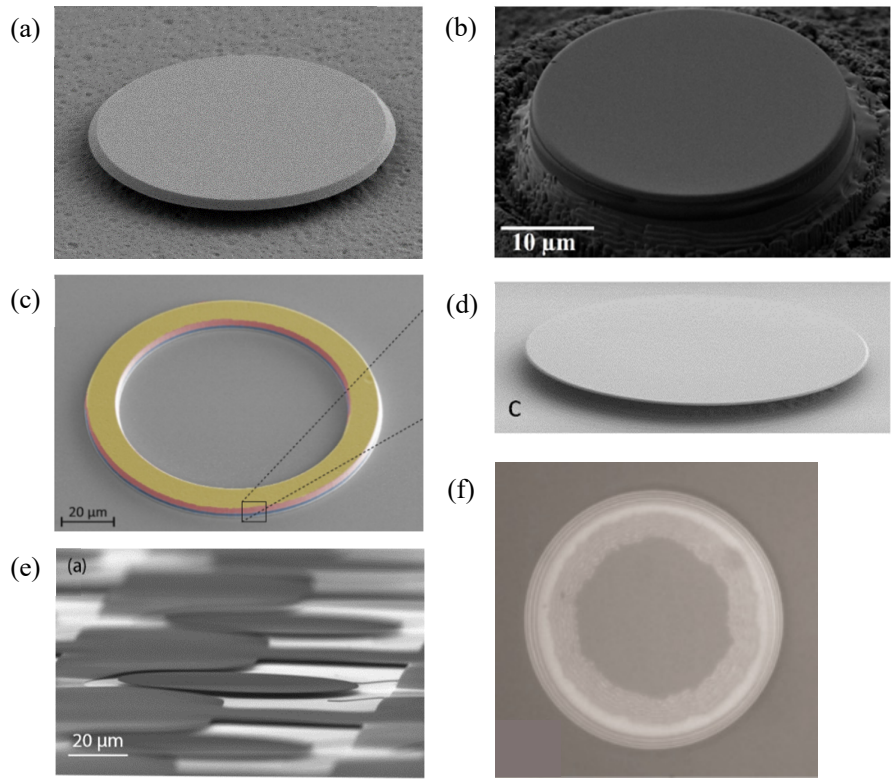

Fig. 6. Examples of dry-etched LN microdisk resonators. Credit: adapted from Refs. [28] (a), [25] (b), [29] (c), [30] (d), [31] (e), [32] (f).

Although it has been shown that an optimized dry etching process itself could create ultrahigh-$Q$ microresonators, such process typically involves a specific set of etching parameters and relies on an ultra-smooth etching mask [17]. Therefore many research groups have developed advanced post-fabrication techniques to further smoothen the etched sidewalls. For example, Y. Cheng's team developed a method of using focused ion beams to smoothen the microdisk sidewalls on a much finer resolution after femtosecond laser machining [19, 25-27]. More recently, K. Buse's team at University of Freiburg showed that mechanical polishing methods can be adopted to improve the microresonator $Q$ factors by more than an order of magnitude [18].

Since the first demonstration of LNOI-based microring resonator in 2007 [24], various types of high-$Q$ microcavities have been realized, a few examples of which are shown in Fig. 6 and Fig. 7. Over the years, we see a steady improvement in $Q$ factors of dry-etched LN microcavities, from $10^4$ in 2012 [33], $10^5$ in 2014 [25, 28], to $10^6$ in 2015 [32], to $10^7$ in 2017 [17-19]. More importantly, we are still over an order of magnitude away from the material absorption limit of LN [21, 22]. We do not see fundamental limitation that prevents dry etched LN microcavities to achieve $Q$-factors comparable with those of SiN [34] or even $SiO_2$ [35]. The dominant loss mechanism in most recent LN microcavities is still surface scattering due to sidewall roughness. However, as the $Q$ factors advance beyond $10^7$, other loss mechanisms, such as absorption losses from cladding and bottom oxides and LN material losses due to crystal damage during ion implantation process, could also become significant. Ways to mitigate these effects, for example by thermal annealing, are yet to be studied.

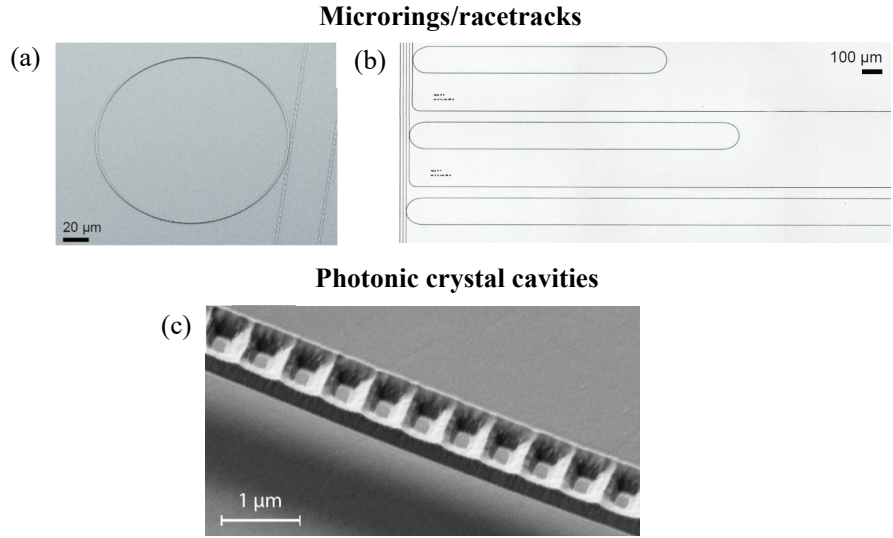

Fig. 7. Examples of waveguide-coupled microring/racetrack resonators (a-b) and photonic crystal cavities (c) in dry etched LNOI platform. Credit: adapted from Refs. [17] (a-b), [36] (c).

The majority of dry etched LN microcavities reported to date have been based on microdisk structures due to the simplicity in device fabrication [18, 19, 25-33, 37-41] (Fig. 6). On the other hand, M. Lončar's team at Harvard University has demonstrated a wide range of integrated LN devices and chip-scale photonic networks, including microring resonators, racetrack resonators and optical waveguides [Fig. 7(a-b)] [15, 17, 20, 42-44]. Due to the slanted sidewalls, producing sub-wavelength features in LN, such as photonic crystal cavities, have been a challenging task for the community [45, 46]. With advanced fabrication techniques, Q. Lin's team at University of Rochester were able to produce photonic crystal cavities with $Q$-factors exceeding $10^5$ in both 1D and 2D configurations [Fig. 7(c)] [36, 47], further demonstrating the versatility of the LNOI platform. Besides microcavities, low-loss optical waveguides based on dry etching LN have also been reported by many groups and used for various applications [48-53].

## 4.2. *Heterogeneous integration*

Despite the tremendous improvement in LN dry etching techniques in recent years, directly structuring the LN device layer still requires nontrivial optimization efforts. Therefore a big part of the LN photonics community has been pursuing the heterogeneous integration approach, which relies on another easy-to-etch material (e.g. Si, SiN) to form ridge waveguides, and exploits the $\chi^{(2)}$ properties of LN in an evanescent fashion. Using this approach, decent microcavity $Q$-factors (close to $10^6$) could be achieved without the need to etch LN.

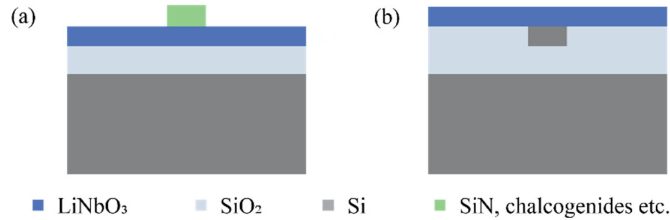

Fig. 8. Schematic illustration of heterogeneous integrated LN photonic devices. (a) Deposition approach. A relatively easy-to-etch material (e.g. SiN) is deposited on top of LNOI substrates and dry etched to form a ridge waveguide. (b) Bonding approach. LN thin film is bonded on top of a pre-fabricated photonic circuit, typically made of Si.

There are two distinct ways of heterogeneous integration: *deposition* approach and *bonding* approach, which are illustrated in Fig. 8. In the deposition approach [Fig. 8(a)], materials that have similar refractive indices as LN are typically used. These materials are deposited on top of the LNOI substrate, and patterned using standard lithography and dry etching techniques. The most popular deposition material is SiN [54-59], which could be deposited and etched relatively straightforwardly. Other deposited materials include chalcogenides (for mid-infrared applications) [60], $TiO_2$ [61] and $Ta_2O_5$ [62]. In the bonding approach [Fig. 8(b)], thin LN films are bonded on top of a pre-fabricated, sometimes larger-scale, photonic circuit, typically made of Si [63-68].

Figure 9 shows a few examples of heterogeneously integrated LN photonic devices using both approaches. Specifically, S. Fathpour's team at CREOL [56-58, 60, 62] and J. Bowers' team at UCSB [14, 54, 59] have done extensive research using the deposition approach. R.M. Reano's team at Ohio State University [63, 64], S. Mookherjea's team at UCSD [65, 66] and A. Safavi-Naeini's team at Stanford University [68] have focused on the integration of thin LN films with Si photonics via bonding approach and its application in electro-optic modulators. The heterogeneous integration approach relies on mature material fabrication technology, therefore could readily produce photonic devices with low optical losses and sub-wavelength features. A. Safavi-Naeini's team was able to fabricate hybrid LN/Si photonic crystal cavities with $Q$-factors as high as $1.2 \times 10^5$ [Fig. 9(d)].

The heterogeneous approaches benefit from their excellent compatibility with existing photonic platforms, and can potentially be integrated into CMOS foundry processes. However, they also share some common drawbacks due to the introduction of alternative materials. The nonlinear interaction strengths in heterogeneous devices are usually compromised since the optical fields only partially overlap with the nonlinear medium, i.e. LN. Moreover, the deposited or bonded materials usually have higher optical absorption losses than LN itself. Although ultrahigh-$Q$ microcavities have been achieved in SiN that is deposited by low-pressure chemical vapor deposition (LPCVD) at high temperatures [34], most SiN/LN hybrid devices to date use PECVD SiN, which is known to possess much higher optical losses. Similarly, Si/LN hybrid devices suffer from carrier absorption and two-/three-photon absorption in Si, and are not suitable for visible applications. In the future, ultrahigh-$Q$ heterogeneous LN microcavities could possibly be realized using LPCVD SiN, which will require the LNOI substrates to sustain the high deposition temperatures.

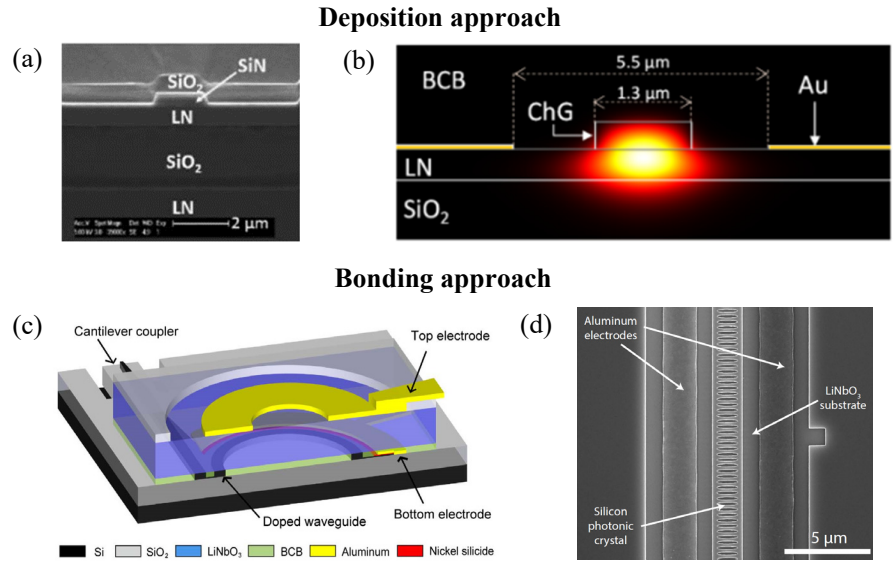

Fig. 9. Examples of thin-film LN photonic devices based on heterogeneous integration. (a-b) Waveguides formed by deposition and etching of SiN (a) and chalcogenides (b). (c-d) Hybrid LN/Si microring resonator (c) and photonic crystal cavity (d) formed by bonding techniques. Credit: adapted from Refs. [54] (a), [60] (b), [64] (c), [68] (d).

## 4.3. *Mechanical polishing*

Besides the aforementioned approaches based on LN thin films, whispering-gallery-mode (WGM) resonators could also be directly realized in bulk LN crystals by mechanically polishing the crystal in a diamond-turning spindle [69]. In fact, this approach has been adopted since early 2000s and applied to many different crystalline materials such as $CaF_2$ and $MgF_2$ [69, 70]. Even today, the mechanical polishing approach still produces LN WGM resonators with the highest $Q$-factors ($> 10^8$), approaching the material limit of LN [21-23, 71]. Figure 10 shows an example of mechanically polished LN WGM resonators reported by L. Maleki's team first at Jet Propulsion Lab and then at OEwaves [21]. The polished WGM resonators typically have a radius of ~ 1 mm and a thickness of 0.05-1 mm [21]. The polishing method benefits from the fact that the bulk of LN crystal remains intact during the process. Therefore the resulting WGM resonators could preserve the intrinsic material properties of LN, e.g. absorption coefficient and nonlinear coefficients. Moreover, crystal engineering prior to the polishing process, such as periodic poling, could also be preserved during the cavity fabrication [21]. However, for the same reason, the dispersion properties of these resonators typically follow those of bulk LN, and could be less flexible for certain nonlinear optical applications such as frequency comb generation.

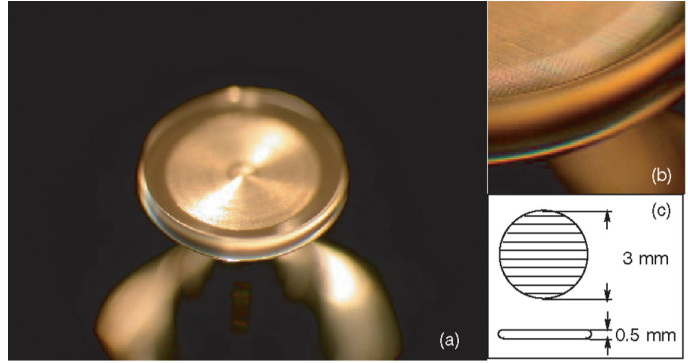

Fig. 10. Example of ultrahigh-$Q$ WGM resonators fabricated by direct mechanical polishing of a LN crystal. Credit: adapted from Ref. [21].

Although the mechanical polishing method is time-consuming and can only produce these resonators one by one, the resulting ultrahigh $Q$-factors make it appealing for high-end precision applications that do not require mass production. Indeed mechanically polished LN resonators have already led to a range of applications including efficient wavelength conversion [21, 71], photon-pair generation [72], single-sideband modulator [73], microwave receiver [74] and quantum microwave-to-photon conversion [23], which will be discussed in Section 6.

## 5. Characterization of LN microcavities

To characterize the optical properties of LN microcavities, various optical coupling schemes have been implemented. For standalone microcavities, evanescent coupling schemes such as tapered fiber and prism coupling, are commonly deployed. In these methods, the coupling strength can be flexibly adjusted and the overall coupling efficiency is high, at the cost of relatively low measurement throughput and stability. Coupling to deformed LN microcavities could also be achieved via free-spacing coupling schemes [41].

A more integrated coupling approach is to directly fabricate the bus optical waveguides in the same device layer as the microcavities, similar to those commonly used in other material platforms [Fig. 11(a)] [17]. The coupling strength could be precisely controlled by lithographically defining the coupling gap *g*. The main challenge in this approach, however arises from the large mismatch between the optical mode sizes of the bus waveguides (< 1 µm) and a typical single-mode optical fiber (~ 10 µm). For end-fire coupling method, fine polishing of the waveguide input/output facets and the use of tapered lensed fibers are usually required. Yet the resulting coupling efficiencies are typically ~ 5 dB/facet without special engineering of the bus waveguide [54, 75]. Most recently, fiber-to-fiber insertion losses of 3.4 dB has been demonstrated by adopting a double-layer inverse taper structure [76]. Since the waveguide dimensions and refractive index contrasts in LNOI platform is similar to other major integrated photonics platforms, we believe fiber-to-chip coupling losses of < 1 dB/facet is not an unrealistic goal for LNOI with further optimizations.

Efficient fiber-to-chip coupling could also be achieved using grating couplers [Fig. 11(b)] [60, 67, 77-79]. The main challenges for implementing grating couplers in LNOI platforms come from the slanted etched sidewalls and the requirement for a partially etched LN layer. The coupling loss currently achieved in etched LN grating couplers are typically 5 – 10 dB per coupler [77, 78]. High-efficiency grating couplers based on the heterogeneous integration approach is in principle easier to achieve, since the infrastructure developed for silicon photonics could be readily used. However, the typical coupling losses achieved to date are still in the 5 – 10 dB per coupler range [60, 67, 79].

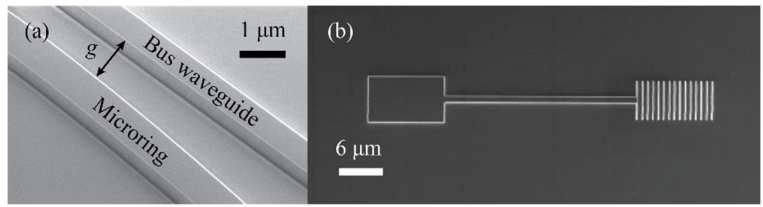

Fig. 11. (a) SEM image showing the coupling region between a microring resonator and a bus waveguide, both fabricated in the same LN device layer. (b) SEM image of a LN grating coupler. Credit: adapted from Ref. [17] (a), [78] (b).

A typical frequency scan of a high-Q LN microresonator at telecom frequencies is displayed in Fig. 12(a), showing a critically coupled resonance with a loaded (intrinsic) *Q* factor of $5 \times 10^6$ ($1 \times 10^7$). We note that for ultrahigh-*Q* (> $10^6$) resonators, a simple laser scan could render rather large measurement errors due to uncertainties in laser frequency calibration. Therefore we recommend that a calibrated laser scan [Fig. 12(a)] and/or a lifetime measurement [Fig. 12(b)] be provided as further confirmation of these ultrahigh $Q$ values. In Fig. 12(a), the input laser light is first modulated at 500 MHz using an external modulator and a precise RF source before being sent to the device. The generated 500-MHz sidebands are therefore projected to the final transmission spectrum and used as a calibration reference. In Fig. 12(b), a cavity ring-down measurement is performed by abruptly switching off the input laser and monitoring the output optical intensity decay. $Q$ factors can subsequently be extracted from the measured cavity lifetime.

Earlier demonstrations of LN microcavities are mostly made in the telecommunication wavelength ranges. More recent results show that $Q$ factors of > $10^7$ could also be achieved in the visible wavelength range, an example of which is shown in [19, 20]. Thin-film LN devices that operate at mid-infrared wavelengths have also been reported in recent years [79].

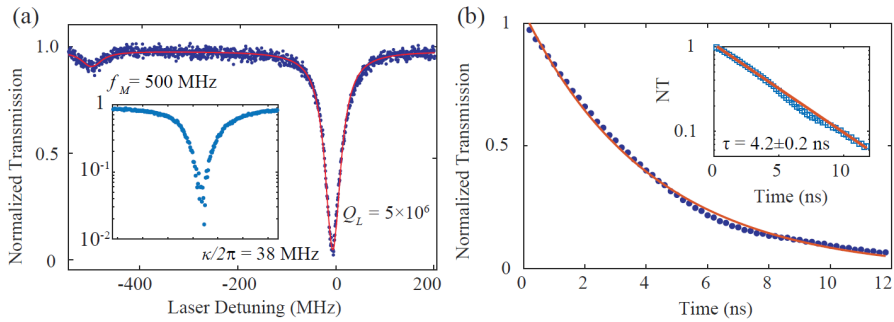

Fig. 12. (a) Transmission spectrum of a typical high-$Q$ LN microcavity. Here a 500-MHz modulated sideband is used as a frequency calibration reference. (b) Cavity ring-down measurement of the same resonance shown in (a), revealing a cavity lifetime of 4.2 ns. Credit: Adapted from Ref. [17].

At elevated optical powers, LN microcavities could experience various nonlinear effects, resulting in deviation from the Lorentzian-shaped static resonance in the low-power region. Apart from thermal-optic effects that are common in almost all materials, LN also experience the photorefractive effect. Quite interestingly, thermal effect induces a red shift of the cavity resonance while the photorefractive effect induces a blue shift, and the two effects occur on vastly different time scales (thermal time constant ~ 10 μs and photorefractive time constant ~ 50 ms). As a result, competition between the two effects could cause cavity resonance to show complex oscillation dynamics at certain optical power levels (Fig. 13), as is studied extensively in Ref. [38], sometimes making it difficult to stabilize the input laser frequency within cavity resonance. We note, however, that the photorefractive effects see quenching behavior at high optical powers, possibly due to elevated local temperatures [36, 42]. In these cases, thermal-optic (red) bistability dominates and cavity resonance becomes stable again.

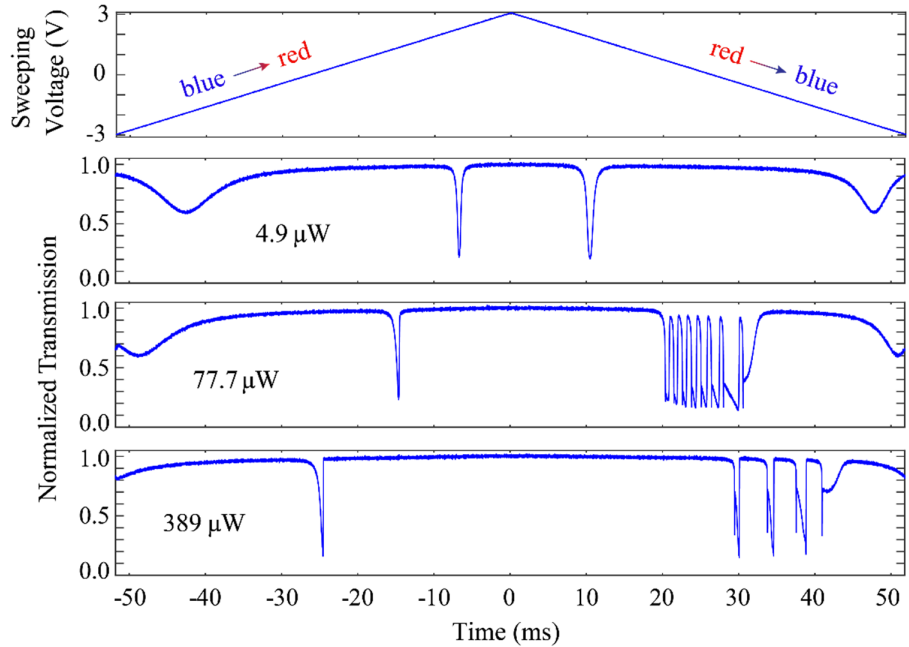

Fig. 13. Forward and backward frequency scans of a high-Q LN microresonator at different power levels. Competition between thermal optic effect and photorefractive effect results in oscillating behaviors when laser is scanned from red side of the resonance towards blue. Credit: Adapted from Ref. [38].

It is well known that photorefractive effect in LN could cause crystal damage and device malfunction above certain threshold optical powers. However, several high-power (> 10 W circulating power) optical measurements in LN microcavities have shown that LNOI platform seems to be more resistant to photorefractive damage than their bulk counterparts [36, 42, 80]. This phenomenon can likely be attributed to two reasons. First, the photorefractive effect in thin-film LN relaxes two to three orders of magnitude faster than that in bulk LN crystals [38]. Second, the refractive index contrast in LNOI waveguides are much larger than that induced by photorefractive effect, making the optical modes largely unaffected by photorefractive effect [14].

## 6. Applications of LN microcavities

Since LN material specializes in its large and convenient $\chi^{(2)}$ coefficient, the majority applications of LN microcavities have been closely associated with $\chi^{(2)}$ processes, namely nonlinear wavelength conversion and electro-optic modulation. Here we provide an overview of current and potential applications of LN microcavities.

### 6.1. *Second harmonic generation*

Because of the strong $\chi^{(2)}$ nonlinearity of LN, an almost immediate phenomenon that can be observed in a high-$Q$ LN microcavity is SHG. For example, when light is coupled into a high-$Q$ optical mode at ~ 1550 nm wavelength, the generated SHG light near red can be readily seen using a visible CCD camera [Fig. 14(a)]. The input-output power relations follow a quadratic behavior, as expected from the physical nature of the second-order nonlinearity [Fig. 14(b)]. Without special engineering, the generated SHG signal is usually quite weak due to momentum and resonance frequency mismatch between the optical modes at fundamental and second harmonic frequencies.

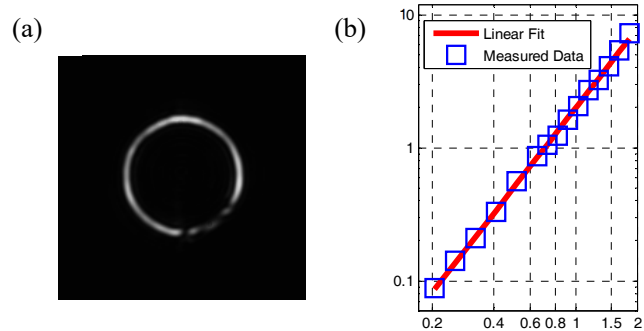

Fig. 14. (a) Top-view visible camera image of a LNOI microresonator when pumped at ~ 1550 nm wavelength. Scattered SHG light can be seen on the circumference of the resonator. (b) Input-output power relationship of the SHG process. Credit: Adapted from Ref. [28].

To achieve phase-matched SHG processes, many different strategies have been proposed and realized, including cyclic phase matching [19, 31, 81], modal phase matching [50, 51, 82], structural phase matching [50, 83] and those based on crystal birefringence [29, 30]. Since SHG process follows a quadratic dependence between input and output optical powers, the normalized conversion efficiency in the low-conversion limit is typically reported, which is defined as $\eta = P_{out}/P_{in}^2$, where $P_{in}$ and $P_{out}$ are input and output powers respectively. We note that the best reported SHG efficiencies in LNOI microresonators to date are still on the order of 100%/W [19, 29-31, 40, 81, 84]. These values are much lower than that already achieved in AlN microresonators (2500%/W) [85] despite that LN's $\chi^{(2)}$ nonlinearity is almost an order of magnitude higher (Table 1). Moreover, SHG in LNOI-based microcavities has not been able to reach the pump-depletion region (high absolute conversion efficiency), possibly due to a combination of relatively low normalized efficiency and resonance instability at high pump powers. The SHG performances in LNOI platform is also lagging behind those based on mechanically polished LN WGM resonators, where a normalized SHG efficiency of ~ 500,000%/W and a total conversion of 9% have been demonstrated [71].

The main factor that limits the conversion efficiencies in LNOI microcavities is the non-ideal nonlinear overlap in these approaches. For example, cyclic phase matching only makes use of LN's nonlinearity at certain locations along the circumference. Phase matching based on crystal birefringence typically uses the $d_{31}$ tensor component due to polarization requirement, which is compromised from the highest $d_{33}$ component. Also, most of the current demonstrations suffer from much lowered collection efficiencies at second harmonic frequencies since the same optical fiber or bus waveguide is used for both pump and collection purposes. This issue could be solved by separately engineering the coupling waveguides at pump and second harmonic wavelengths, similar to that used in [85, 86].

Quasi-phase matching (QPM) based on periodically poled lithium niobate (PPLN), could become a promising candidate for achieving ultrahigh-efficiency wavelength conversion in LNOI microresonators. Traditionally, PPLN crystals and ion-diffused waveguides have been widely used in nonlinear optics, where the $z$ axis of LN crystal is periodically inverted to compensate for the phase mismatch during the nonlinear process [87]. Incorporating PPLN with high-$Q$ microcavities, where the optical intensity could be resonantly enhanced by many orders of magnitude, could dramatically increase the device efficiencies. The main challenge for achieving PPLN in thin LN films comes from the requirement of a much smaller poling period (~ 5 µm) compared with that in traditional PPLN (~ 20 µm) due to a much stronger geometric dispersion [40, 54, 80, 84, 88]. Recent years have seen a lot of progress in realizing high-quality periodic poling in LNOI platform (Fig. 15). In LNOI waveguides, high-quality periodic poling with periods down to 4 µm and normalized efficiencies more than an order of magnitude higher (2600%/W-cm$^2$) than ion-diffused PPLN waveguides have been experimentally realized [Fig. 15(b)] [80]. High-$Q$ PPLN microcavities have also been realized at a poling period of 23 μm, whose conversion efficiency is currently limited since the $d_{22}$ coefficient is used [Fig. 15(c)] [40]. With further co-optimization of the periodic poling process and the LN microcavities, we expect the SHG conversion efficiencies of LNOI microcavities to see significant further increase.

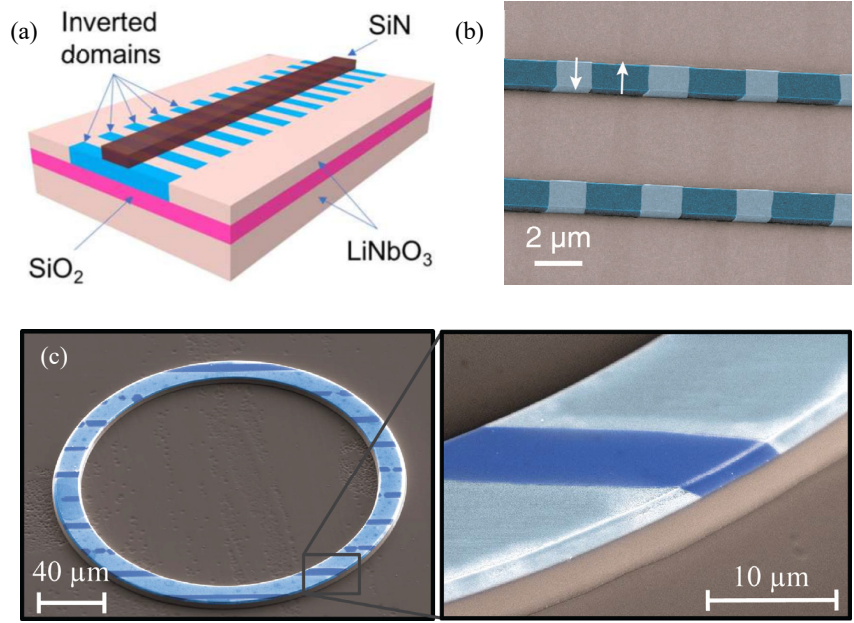

Fig. 15. Examples of PPLN realizations in LNOI platform. Credit: Adapted from Ref. [54] (a), [80] (b), [40] (c).

### 6.2. *Electro-optic modulation*

When an external electric field is applied to a LN microcavity, the change of refractive index modifies the effective optical path length of the resonator, resulting in a resonance frequency shift. This effect could be of interest for a range of applications including tunable photonic devices, electro-optic modulators and compact optical switches. The electro-optic efficiencies in thin-film LN platform are usually much higher than those in conventional ion-diffused LN devices, since the metal contacts can be placed much closer to the highly confined optical waveguides without introducing detrimental optical losses [15, 27, 32, 52, 55, 57, 58, 63-65, 68, 75, 79, 89, 90].

There are three major device configurations for electro-optically tunable LN microcavities, as are shown in Fig. 16. In the early work from P. Günter's team at ETH Zurich, a z-cut LN and top/bottom electrode architecture is used, where the measured electro-optic efficiency for TM light is ~ 1 pm/V [Fig. 16(a)] [24]. The relatively low efficiency here is partly due to the lowered $r_{33}$ values (~ 50% of the native value) in their LN thin films, which were not annealed at a high enough temperature [24]. Although the top/bottom-electrode configuration is relatively easy in device fabrication, it has not been widely adopted since the first demonstration in 2007 [24, 32]. This is likely due to the inconvenience in electronic testing, especially at high frequencies.

The second approach uses an x-cut LN configuration, with planar metal electrodes placed side by side with the optical waveguides [Fig. 16(b)] [15, 27, 58, 60, 68]. The electrode layout in an x-cut configuration requires special engineering since light in a WGM resonator travels along different crystal directions on the $y$-$z$ plane. One design from M. Lončar's team at Harvard University that optimizes the electro-optic overlap for TE polarization is shown in Fig. 16(b), where the electrical fields on the two racetrack arms are aligned to the same direction (+$z$) so that the modulations on the two arms add up constructively. Although this approach requires two metal layers and corresponding vias connecting them, it leads to a high electro-optic efficiency of 7 pm/V and could readily support high-frequency (> 100 GHz) RF delivery due to the small resistance and capacitance in the system [15]. Indeed, electro-optic bandwidths of > 30 GHz and open-eye data transmission at up to 40 Gbit/s has been demonstrated using this approach. This planar-electrode configuration also applies to heterogeneous approaches and photonic crystal cavities, as are demonstrated in [58, 60, 68].

For bonding-type heterogeneous integration approach, the underlining Si optical waveguides can also be used directly as an electrode [Fig. 9(c)] [63, 64]. Since the electrodes are in direct contact with the electro-optic material (i.e. LN), a record-high electro-optic efficiency of 12.5 pm/V has been reported by R.M. Reano's team at Ohio State University [63]. However, the use of doped Si as both electrodes and optical waveguides induces a large parasitic resistance as well as significant optical losses to the system. As a result, the 12.5-pm/V measurement is only performed at DC frequency and the later demonstration at 5-GHz electro-optic bandwidth is based on a 3.3-pm/V device using TE polarization [64].

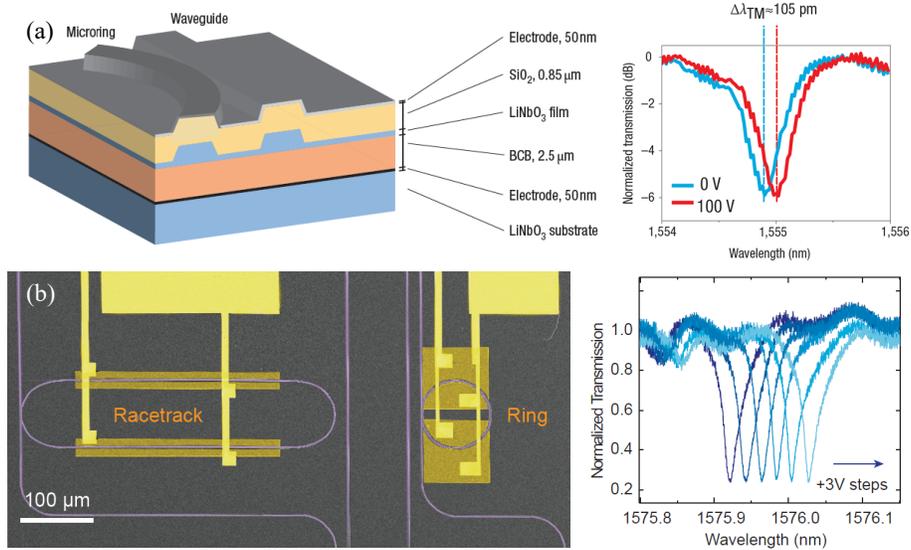

Fig. 16. Examples of electro-optically tunable LN microresonators. (a) z-cut LN microresonator with a top/bottom electrode structure, showing a tuning efficiency of 1 pm/V. (b) x-cut LN microresonators with in-plane metal contacts, showing a tuning efficiency of 7 pm/V. Credit: Adapted from Ref. [24] (a), [15] (b).

The electro-optic bandwidths of these LN microresonators depend on two major factors, namely the electrical circuit RC constant and the cavity photon lifetime. For device configurations that use metal electrodes [e.g. Fig. 16(b)], the RC bandwidths are usually very high since the on-chip metal resistance is much smaller than the typical driving circuitry impedance (50 $\Omega$) and the end capacitance is also small thanks to the tiny device footprint. Instead, the electro-optic bandwidths are often limited by the cavity photon lifetime, since light traveling inside a high-$Q$ microcavity cannot respond to external electric field fast enough. For example, in order to achieve a 40-GHz electro-optic bandwidth at telecommunication wavelengths, the cavity $Q$ factor is required to be less than 5,000, which in turn increases the required switching voltage for a full on-off swing. This voltage-bandwidth trade-off, together with the sensitivity of cavity resonance to environmental fluctuations, could ultimately limit the applications of LN microresonators in telecommunications. On the other hand, ultrahigh electro-optic bandwidths (> 100 GHz) and low drive voltage (< 1.5 V) have been realized in the thin-film LN platform, using a Mach-Zehnder interferometer (MZI) configuration [65, 75, 90].

Despite the voltage-bandwidth trade-off, electro-optically active LN microcavities could be especially appealing for many assignment-specific applications. For example, many microwave photonics and quantum photonics applications require single-sideband modulators, which cannot be achieved by a simple MZI modulator without losing optical power. Such functionality could be realized using a doubly resonant LN microcavity, for example the coupled-resonator system shown in Fig. 17(a), where the frequency difference between the two optical resonances is designed to match with the desired modulation frequency [44]. Due to the asymmetry in the device spectral response, only one of the two modulation sidebands is matched with an optical resonance, while the other sideband is greatly suppressed [Fig. 17(b)] [91]. Similar functionality has also been realized using TE and TM modes of a polished LN WGM resonator [Fig. 17(c-d)] [73]. Such devices can also be used as a tunable narrowband microwave receiver [74] and a photonic buffer that supports on-demand photon storage and retrieval [44].

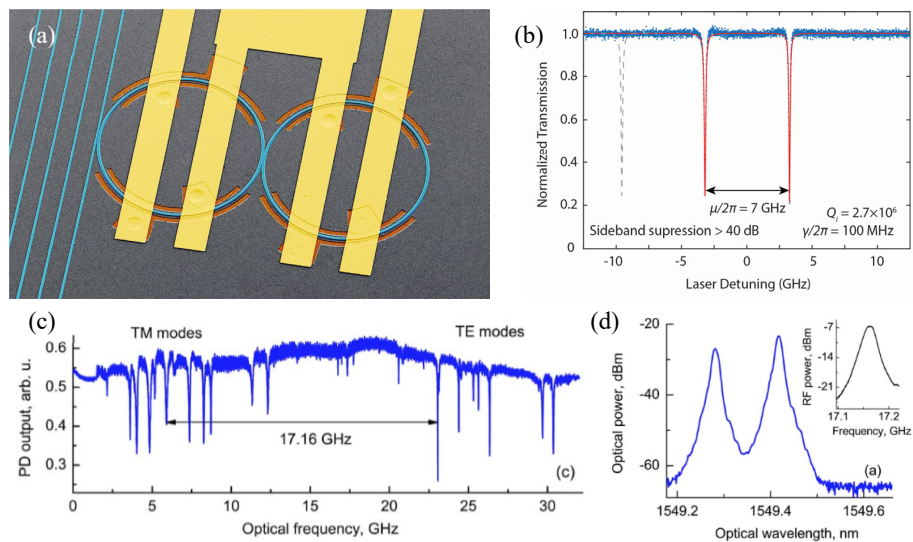

Fig. 17. Examples of doubly resonant LN microresonators for single-sideband modulation. (a) SEM image of a thin-film LN coupled-resonator system. (b) Optical transmission spectrum of the coupled resonator showing an asymmetric spectral response, which is used to suppress the modulation sideband on the left. (c) Optical transmission spectrum of a polished LN resonator showing TE and TM modes separated by 17.16 GHz. (d) Single-sideband modulation output spectrum of the device shown in (c). Credit: Adapted from Ref. [44] (a), [91] (b), [73] (c-d).

### 6.3. *Frequency comb generation*

Optical frequency combs are excellent broadband coherent light sources and precise spectral rulers [92]. In recent years, Kerr frequency comb generation based on high-*Q* microcavities has emerged as a particularly promising candidate for a range of applications including optical clocks, pulse shaping, spectroscopy, telecommunications, light detection and ranging (LiDAR) and quantum information processing [93]. Although Kerr combs have been realized in various material platforms, realizing frequency combs in LN is still of great interest to the comb community due to the existence of a strong $\chi^{(2)}$ nonlinearity, which could be exploited to assist the comb generation process and to achieve extra device functionalities.

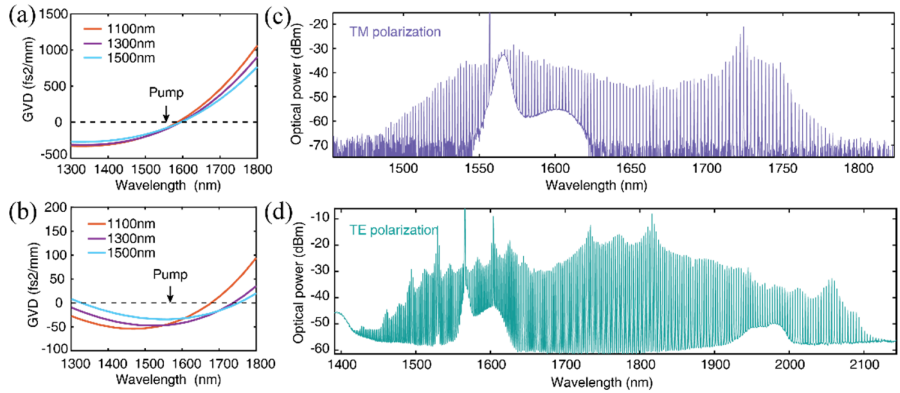

Fig. 18. (a-b) Numerically simulated group-velocity dispersions (GVD) for LNOI waveguides with different top widths, achieving anomalous dispersions (GVD < 0) for both TM (a) and TE modes. (c-d) Generated Kerr frequency comb spectra for TM (c) and TE (d) polarizations. Credit: Adapted from Ref. [42].

Interestingly, LN does possess a decent $\chi^{(3)}$ coefficient of $1.6 \times 10^{-21}$ m$^2$/V$^2$, similar to that of SiN which is commonly used for Kerr comb generation. Apart from the good $\chi^{(3)}$ nonlinearity, the LN microresonator is also required to have a high *Q* factor and anomalous dispersion in order for the optical parametric oscillation (OPO) process to take place. While mechanically polished LN WGM resonators could achieve ultra-high *Q* factors of > $10^8$, their dispersion properties are predetermined by the bulk material properties (usually normal dispersion) and cannot be engineered. In contrast, LNOI platform is ideally suited for dispersion engineering due to the strong light confinement [39]. Indeed, anomalous dispersion can be achieved for both TE and TM polarizations in LNOI microrings without exotic design efforts [Fig. 18(a-b)] [42]. When pumping these high-$Q$ resonators at optical powers of ~ 100 mW, frequency combs were generated for both polarizations, spanning spectral ranges as wide as 700 nm [Fig. 18(c-d)] [42]. More recently, single- and multiple-soliton states have been successfully generated by temporal scanning methods, further demonstrating the potential of LNOI microring resonators for frequency comb applications [94].

Perhaps what's more important than the generation of Kerr combs in LN itself, is to combine LN combs with its $\chi^{(2)}$ nonlinearity. For example, by cascading a second microring resonator as a tunable filter, the generated Kerr comb spectrum can be further shaped and manipulated on the same chip (Fig. 19) [42]. Thanks to the ultrahigh electro-optic tuning bandwidth available in LN, such level of integration could enable a range of microcomb applications that require fast intensity/phase modulation, such as dense-wavelength-division multiplexing (DWDM) systems for future ultra-broadband optical fiber communication networks [95].

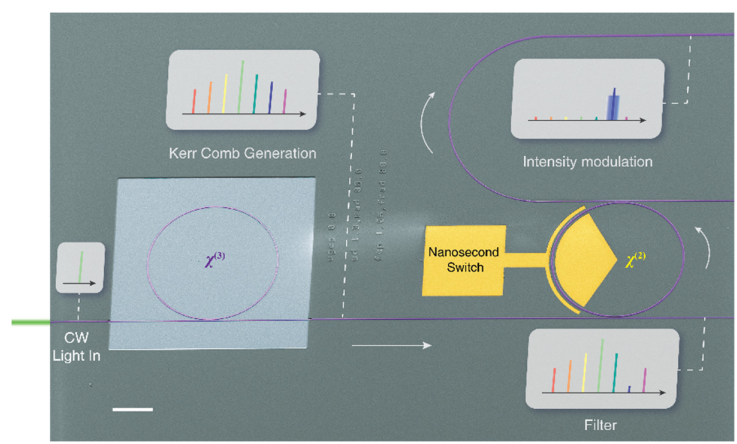

Fig. 19. False-colored SEM image showing a fabricated LN nanophotonic circuit that consists of a microresonator frequency comb generator ($\chi^{(3)}$) and an electro-optically tunable add-drop filter ($\chi^{(2)}$). Credit: Adapted from Ref. [42].

Remarkably, without the need for $\chi^{(3)}$ nonlinearity, LN's strong $\chi^{(2)}$ effects can also be used to directly generate a broadband frequency comb in LN microcavities. It is well known that electro-optic frequency combs can be generated by passing a continuous-wave laser through a sequence of phase and amplitude modulators [96]. However, such devices usually generate only tens of lines and spanning only a few nanometers [96]. By embedding the electro-optic frequency comb generation process into a high-$Q$ LN microcavity, such process can be dramatically enhanced, resulting in much broader frequency combs (> 90 nm) (Fig. 20) [43]. Compared with Kerr combs, electro-optic combs can be generated at arbitrary optical powers (threshold-less), can be flexibly controlled using microwaves, and usually have good noise properties due to the low-noise nature of microwave sources. Further optimizing the resonator $Q$ factors and dispersion properties could potentially result in octave-spanning electro-optic combs.

Other than the electro-optic effect, LN's $\chi^{(2)}$ effects could also be explored in the nonlinear optics perspective for low-power and ultra-broadband frequency comb generation. For example, the threshold power of $\chi^{(3)}$ frequency comb generation in LN microcavities could potentially be dramatically lowered by leveraging the giant $\chi^{(2)}$-induced effective $\chi^{(3)}$ nonlinearity in a quasi-phase-matched waveguide [97]. Frequency combs could also be generated directly from $\chi^{(2)}$ frequency down conversion [98]. By exploring various frequency conversion processes, including SHG, third-harmonic generation and sum-/difference-frequency generation (SFG/DFG), frequency combs that spans from visible to mid-infrared could possibly be realized in LN microcavities in the future.

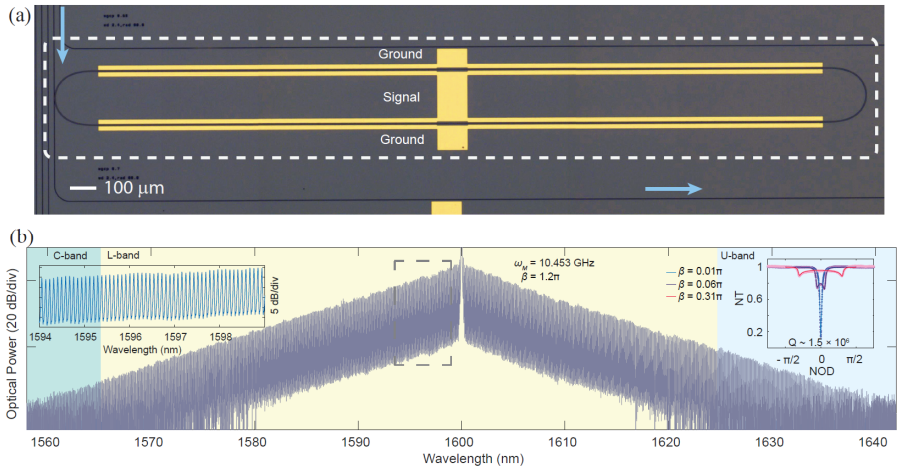

Fig. 20. (a) False-colored SEM image of a fabricated LN racetrack resonator for electro-optic comb generation. (b) Measured output spectrum of the generated electro-optic comb with > 80 nm bandwidth and more than 900 comb lines. Credit: Adapted from Ref. [43].

**6.4.** *Quantum photonics*

Since the $\chi^{(2)}$ and $\chi^{(3)}$ nonlinearities intrinsically preserve the quantum coherence of the photons involved in the processes, many of the abovementioned applications of LN microcavities can be readily extended to single-photon levels and used for quantum applications. The ultralow optical loss achievable in LNOI photonics is particularly appealing for quantum photonic circuits since quantum information cannot be cloned. In this section, we provide an overview of important applications of LN microcavities in quantum photonics.

When pumping the SHG devices discussed in Section 6.1 at the second-harmonic frequencies, photon pairs can be generated via spontaneous parametric down conversion (SPDC). These photon pairs are intrinsically entangled with each other, therefore are important photon sources for many quantum optics applications. Photon pair generation with high SPDC rates and good quantum coherence has been demonstrated in high-$Q$ LN microcavities [31, 99]. Similar wavelength-conversion devices are also promising for converting single photons from one wavelength to another with near-unity efficiency and high quantum fidelity, potentially acting as a quantum interface that connects various quantum platforms (e.g. quantum dots, diamond color centers and atomic systems) and popular photon transmission channels (e.g. optical fibers).

The electro-optic devices discussed in Section 6.2 can also be used for quantum applications, operating as a quantum microwave-to-optical converter. This is highly relevant for superconducting qubit based quantum systems, since microwave superconducting qubits cannot be transmitted outside the dilution fridges (~ mK temperature) due to overwhelming thermal noises. By converting the microwave qubits into high-energy optical photons via LN's electro-optic effect, quantum information can be transmitted without decoherence over long distances. In order to achieve near-unity microwave-to-optical conversion, apart from a high optical $Q$ factor and a good electro-optic overlap, the system is also required to have a high-$Q$ microwave cavity. In Ref. [23], a maximum microwave-to-optical photon-number conversion efficiency of 0.1% was demonstrated using a mechanically polished LN resonator (Fig. 21). Although the system benefits from an ultrahigh optical $Q$ factor of ~ $10^8$, the conversion efficiency is limited by the low microwave $Q$ factor of ~ 200 of the open copper cavity used, as well as the relatively low electro-optic overlap of the bulk LN configuration. Theoretical calculations have suggested that, by combining a microwave cavity with decent $Q$ factors (~ $10^4$, commonly achieved in superconducting microwave circuits) with state-of-the-art LN electro-optic microcavities, near-unity conversion can be achieved at µW level of optical pump powers [100, 101].

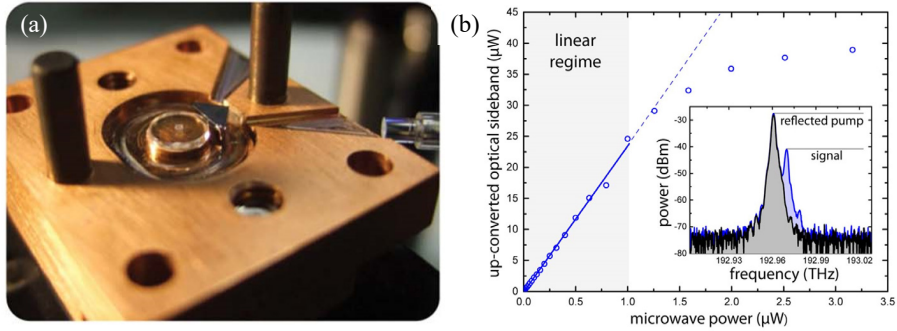

Fig. 21. (a) Photograph of a mechanically polished LN resonator embedded in the bottom part of an open copper microwave cavity. (b) Power of the up-converted optical sideband as a function of the microwave power sent to the cavity, showing a photon-number conversion efficiency of ~ 0.1%. Credit: Adapted from Ref. [23].

## 7. Outlook

Although $Q$ factors of LNOI microcavities today are already among the top tier of various integrated photonic platforms, we believe the development of LNOI platform is still in its very early age, and many more important milestones are yet to come. Apart from the applications that already discussed above, here we provide a few other promising example applications of LN microcavities that are not fully explored yet.

It is well known that various doping in LN could significantly modify the crystal properties and similar techniques could be adopted in thin LN films. For example, MgO:LN [102] and $ZrO_2$:LN [103] are known to suppress the photorefractive damage in LN crystals. Er:LN could potentially be used to realize on-chip optical amplifiers in LNOI platform [104]. Tm:LN has been reported to have good quantum properties, promising for achieving a quantum memory in LN [105]. Studies on the doping conditions and effects, however are still very preliminary in the LNOI platform, and are subject to future research.

Currently most studies of LN microcavities have been carried out in the telecommunications bands, with some exceptions in visible wavelengths, while LN's transparency window spans from blue to mid-infrared. Combining the low-optical losses, electro-optic and nonlinear optical properties, LN photonics outside the traditional telecommunications bands could find a wide range of applications in bioimaging, spectroscopy and quantum technologies.

LN also exhibits interesting mechanical properties due to its strong piezoelectric responses. Indeed LN has been one of the most important surface-acoustic wave (SAW) materials for microwave filters [3]. Combining the ultrahigh-$Q$ factors and strong piezoelectricity of LN microcavities could potentially lead to new optomechanics phenomena and applications.

Finally, the active electro-optic modulation available in LN microcavities provides an extra degree of freedom in integrated photonics, and therefore could be used to explore various new physics. For example, electro-optic modulation could provide an effective magnetic field to an optical system, which can be used to build non-reciprocal devices (e.g. integrated isolators) [106] and to explore topological photonics [107].